\documentclass[aps,showpacs,pra,twocolumn]{revtex4}
\usepackage{epsfig,amssymb,amscd,amsmath}
\usepackage{subfigure}
\usepackage{epsfig}
\usepackage{graphicx}

\newcommand{\Tr}{{\mbox{Tr}}}
\newcommand{\Id}{1\hspace{-0.56ex}{\rm I}}
\newcommand{\be}{\begin{equation}}
\newcommand{\ee}{\end{equation}}
\newcommand{\ket}[1]{ | \, #1  \rangle}
\newcommand{\bra}[1]{ \langle #1 \,  |}
\newcommand{\bk}[2]{ \langle #1\, |\, #2 \rangle }
\newcommand{\bea}{\begin{eqnarray} }
\newcommand{\eea}{\end{eqnarray} }
\newcommand{\non}{\nonumber}

\begin{document}
\bibliographystyle{prsty}

\title{Entanglement, purity and energy: Two qubits {\it vs} Two modes}
\author{Derek McHugh$^{1,5}$, M\'ario Ziman$^{1,2,3}$ and Vladim\'\i r Bu\v zek$^{1,3,4}$}
\affiliation{ $^1$Research Center for Quantum Information, Slovak
Academy of Sciences, D\'ubravsk\'a cesta 9, 84511 Bratislava,
Slovakia
\\
$^2$Faculty of Informatics, Masaryk University, Botanick\'a 68a,
60200
Brno, Czech Republic\\
$^3$Quniverse, L\'\i\v s\v cie \'udolie 116, 841 04 Bratislava, Slovakia\\
$^4$Abteilung f\"{u}r Quantenphysik, Universit\"at Ulm, 89069 Ulm,
Germany\\
$^5$Department of Mathematical Physics, National University of
Ireland Maynooth, Maynooth, Co. Kildare, Ireland}

\received{\today}

\begin{abstract}
We study the relationship between the entanglement, mixedness and
energy of two-qubit and two-mode Gaussian quantum states. We
parametrize the set of allowed states of these two fundamentally
different physical systems using measures of entanglement,
mixedness and energy that allow us to compare and contrast the two
systems using a phase diagram. This phase diagram enables one to
clearly identify not only the physically allowed states, but the
set of states connected under an arbitrary quantum operation. We
pay particular attention to the maximally entangled mixed states
(MEMS) of each system. Following this we investigate how efficiently one
may transfer entanglement from two-mode to two-qubit states.
\end{abstract}
%\vskip 0.1cm
%\narrowtext
\maketitle

\section{Introduction}
In this paper we present a parametrization of the states of two
qubits and the Gaussian states of a two-mode continuous variable
system in terms of entanglement, purity and energy. This allows
one to compare these two fundamentally different physical systems
using a entanglement-purity-energy (EPE) phase diagram. The
relationship between these three properties gives us some insight
into the set of allowed states in both systems and, in particular,
the distinction between separable and entangled states, both pure
and mixed. Also there has been considerable interest in recent
years in the maximally entangled mixed states (MEMS) of two qubits
and two modes. Here we re-examine this type of state and its
dependence on the energy in the system.

Entanglement is the resource used in many quantum information
protocols including quantum teleportation, quantum cryptography
and quantum communication. In the simplest form, these protocols
ideally require maximally entangled pure states. It is highly
likely though that in any real implementation the states will be
mixed to some degree. It is then desirable to know how this
mixedness limits the amount of entanglement.

Qubits have been the mostly studied basic information unit in
quantum information theory. As a two-level quantum system they
have been experimentally realised in many different physical
set-ups. Two qubits provide the simplest discrete quantum system
where non-trivial properties of entanglement can be studied. The
entanglement can be unambiguously quantified using different
measures and the maximally entangled pure states are the so-called
Bell pairs. For mixed states, those states which maximise the
entanglement for a given degree of purity are the MEMS states
introduced by Ishizaka and Hiroshima \cite{SITH}. They were
studied in more depth by Verstraete {\it et al.} in Ref. \cite{FVKA} who
derived the unitary operation which must be applied in order to
maximise the entanglement of a state $\rho$. An Ansatz for the
MEMS, which is seen to be locally equivalent to a diagonal state
transformed under one of the aforementioned unitary operations,
was presented in Ref. \cite{VMKG}. Different measures of entanglement
do not have the same dependence on a given measure of mixedness.
Thus the MEMS depend on which measures are chosen and this fact is
investigated in detail in Ref. \cite{VMKG}.

The previous studies of maximally entangled mixed states use
entanglement and mixedness to characterize the set of two-qubit
states. Indeed, in Ref. \cite{MZVB} the authors use a concurrence-purity (CP) phase diagram to graphically present the set of physically allowed states. A physical realization of two qubits could be two
two-level atoms or two polarized photons where the $\ket{0}$ and
$\ket{1}$ qubit levels are the ground and excited atomic levels in
the former or vertical and horizontal polarization states in the
latter case. It is the case that the energy in the system has a direct
influence on the entanglement and mixedness properties. For
instance, if we know there are zero or two excitations in the
system composed of two atoms we can tell without ambiguity that the two-qubit is both
pure and separable. For other energies the connection between
entanglement, mixedness and energy is not so clear. The first aim
of this paper is to explore this relationship with particular
emphasis on ``frontier'' states - those states that lie on the
edge of the physically allowed states.

We will also investigate the relationship between entanglement,
mixedness and energy for two-mode Gaussian states. Interest in the
information processing abilities of continuous variable quantum
systems has grown dramatically in recent years due mainly to the
experimental benefits of generating and measuring entangled states
compared to a discrete quantum system. The simplest system one can
then study is that of two modes of, say, an electromagnetic field.
Also, by restricting to Gaussian states, those states with a
Gaussian Wigner phase space representation, there exists solid
measures of entanglement when the energy in the system is finite.
The idea of a Bell state in this system is not clearly defined as
it is possible in theory to have arbitrarily large entangled
states limited only by the ability to squeeze states.

Extremally entangled states of two-mode Gaussian states were
studied in Refs. \cite{seradill1,seradill} where it was demonstrated
that by fixing not only the global purity, but also the purities
of the reduced density matrices of both modes, there exist
maximally and minimally entangled states. In what follows we will
simply fix the total energy in the system and then look for the
physically allowed states at this energy, including those that are
maximally entangled. The EPE phase diagram offers a nice visual
aide to the allowed states. The distinction between separable and
entangled states is clear to see and, while maximally entangled
states coincide with the previous work, in this picture the
minimally entangled states are just the separable states.

The rest of the paper is organised as follows. In section
\ref{2qubit} we cover the case of two qubits in some detail with
particular regard to the extremal states. Section \ref{2modes}
covers the two-mode Gaussian states where we find some
similarities and differences with previous efforts to describe
these states. In section \ref{CPE2q2m} we look
at how entanglement can be transferred from two-mode to two-qubit states.
We conclude in section \ref{conc} with some closing remarks.

\section{Two-qubit states}
\label{2qubit}
 
% Figure 1 
%  \begin{figure*}[floatfix]
%  \begin{center}
%  \setlength{\unitlength}{1cm}
%  %\includegraphics[width=8cm\textwidth,keepaspectratio]{./paper_CPMEMS.png}
%  \includegraphics[width=8cm]{./paper_CPmemsW.eps}
%  \end{center}
%  \caption{The concurrence {\it vs} purity phase diagram of two
%  qubits. The lower curve, $C_W$, indicates the Werner states, while the upper curve corresponds to the MEMS. All quantities are dimensionless.} \label{CPpd}
%  \end{figure*}

Two qubit states are completely specified by 15 parameters and as such, it
is difficult to find an appropriate parametrization in which to
view the state space. Therefore we want to reduce the number of
parameters involved while simultaneously retaining as much
non-trivial information about the states as possible.

We choose to use the entanglement, purity and energy. First we
need to define the quantities we will use to quantify the
entanglement, mixedness and energy for qubits. For entanglement we
choose to employ the concurrence. This is an easily calculable
measure for the entanglement of two qubits in the state $\rho$
given by the explicit formula, \be
C(\rho)=2\max\{\mu_j\}-\sum_j\mu_j\;\;,\ee where $\{\mu_j\}$ are
the square roots of the eigenvalues of the matrix $R=\rho
(\sigma_y\otimes\sigma_y) \rho^* (\sigma_y\otimes\sigma_y)$ and
$\rho^*$ is the complex conjugate of the density matrix $\rho$.
$C(\rho)$ ranges from zero for separable states to one for
maximally entangled Bell states. The tangle, \be \tau(\rho)=[C(\rho)]^2\;\;,\ee and
the entanglement of formation, \be EoF(\rho)=h(x_+)+h(x_-)\;\;,\ee where $x_{\pm}=(1\pm\sqrt{1-[C(\rho)]^2})/2$ and, \be h(x)=-x\log_2 x\;\;\ee and the concurrence are equivalent measures of entanglement. The negativity, \be N(\rho)=
\frac{||\rho^{\Gamma}||-1}{2}\;\;,\ee and the logarithmic negativity, 
\be L_N(\rho)=\log_2||\rho^{\Gamma}||\;\;,\ee where 
$||A||={\mbox Tr}\sqrt{AA^{\dagger}}$ and
$\rho^{\Gamma}$ is the partial transpose of $\rho$ with respect to one or other of the subsystems, are two measures of entanglement we could also have used here. The ordering of states is changed
depending on the choice of entanglement measures \cite{eisplen,vers,mario}. In Ref. \cite{FVKA} the authors investigate in detail how the maximally entangled mixed states depend on the choice of measures for both entanglement and mixedness. For instance, when purity is chosen to measure the mixedness and concurrence the entanglement, the MEMS are those states defined in Eq. (\ref{MEMS}) whereas if the negativity is used to measure entanglement, the Werner states, \be \rho_W=r\ket{\Psi}\bra{\Psi}+\frac{(1-r)}{4}\mathbb{I}\;\;,\label{werner}\ee where $\ket{\Psi}$ is a maximally entangled state and $\mathbb{I}$ is the identity, achieve the highest degree of entanglement for a given amount of purity. Here we are interested in what parameters we can use to view the set of physically allowed states of a two-qubit system in such a way as to gain as much information as possible. We take generally the entanglement, the mixedness and the energy, while the measures we choose to quantify these properties are not so important. It only affects the shape of 
 the entanglement-energy-purity diagram as can seen when comparing Fig. 3 to Fig. 4(a).
%the entanglement-energy-purity diagram as can be seen when comparing Fig. \ref{outlineneg} to Fig. \ref{outline}.

To quantify the mixedness of a state $\rho$ we will use the
purity, \be P(\rho)=\Tr[\rho^2]\;\;.\ee In general, for
$d$-dimensional systems $P(\rho)$ ranges from $1/d$ for completely
mixed states to 1 for pure states. It is closely related to the
linear entropy measure of mixedness, $S_L(\rho)= \frac{d}{d-1}
(1-P(\rho))$, $d>1$, which ranges from zero for pure states to one for the maximally mixed $d$-dimensional state.

The energy of two two-level atoms with resonance frequency
$\omega_0$ is defined here as the expectation value of the
Hamiltonian, $H_{atom}= \hbar\omega_0\sigma_z^{(1)}+
\hbar\omega_0\sigma_z^{(2)}$, where $\sigma_z$ is the Pauli matrix
satisfying $\sigma_z\ket{j}= (-1)^{j+1}\frac{1}{2}\ket{j}$, for
the ground and excited atomic states $\{\ket{0},\ket{1}\}$. We set
$\hbar$ and $\omega_0$ to 1 leaving $H=\sigma_z^{(1)}+
\sigma_z^{(2)}$. We will use the average excitation number in the
system, \be E(\rho)=1+\Tr [H\rho]\;\;.\ee $E(\rho)$
ranges from zero (for $\rho=\ket{00}\bra{00}$) to two
($\rho=\ket{11}\bra{11}$) for the physically allowed two-qubit
states.

%In this paper we will study the relationship between these three
%quantities. As a visual aide, we plot the set of physically
%allowed states on a concurrence-purity-energy phase diagram and
%determine the important extremal point on this diagram.

A concurrence and purity phase diagram is one method which tells
us much about the set of physically allowed states. This phase
diagram is shown in Fig. 1. It is bound by the maximally
%diagram is shown in Fig. \ref{CPpd}. It is bound by the maximally
entangled mixed states defined later in Eq. (\ref{MEMS}). Also shown in this figure are the Werner states from Eq. (\ref{werner}). We now add
another parameter the energy of the states. The relationship
between the energy, purity and entanglement is already known for
certain extreme two-qubit states. These are the pure states, the
separable states and the maximally entangled mixed states
mentioned above. The entanglement-purity-energy (EPE) phase diagram
for these states is shown in Fig. 2. Each point in these
%for these states is shown in Fig. \ref{figall}. Each point in these
diagrams represents the set of states, \be \{\rho:E(\rho)=E,
C(\rho)=C, P(\rho)=P \} \;\;.\ee We first go through explicitly
the boundaries of Fig. 2 which are more clearly shown in Fig. 4(a).
%the boundaries of Fig. \ref{figall} which are more clearly shown in Fig. \ref{outline}.

\subsection{Pure States}

For the pure states, $P=1$, we have, \be
\ket{\Psi}=a\ket{\alpha_1}\ket{\beta_1}+
b\ket{\alpha_2}\ket{\beta_2}\;\;, \ee by the Schmidt
decomposition, where $a,b\in\mathbb{R}$, $0\le a,b\le 1$ with
$a^2+b^2=1$, and $\bk{\alpha_j}{\alpha_k}= \delta_{jk}=
\bk{\beta_j}{\beta_k}$. The concurrence for $\ket{\Psi}$ is
$C(\ket{\Psi} \bra{\Psi})=2ab$. The maximum/minimum energy of
$\ket{\Psi}$ for a given concurrence is easily seen to be
$E^{max/min}=2b^2$. Thus for a concurrence $C$, the pure
states on the EPE diagram are contained in a circle centered on
$E=1,C=0,P=1$ with radius 1, $(E-1)^2+C^2\le 1$. This is the
blue curve in Fig. 4 (curve satisfying $P=1$).
%blue curve in Fig. \ref{figoutline} (curve satisfying $P=1$).

% % Figure 2
%  \begin{figure*}[floatfix]
%  \centering \setlength{\unitlength}{1cm}
%  %\includegraphics[width=8cm]{./paper_allstates1.png}
%  \includegraphics[width=8cm]{./paper_allstates1.eps}
%  \caption{(Color online) Numerically determined entanglement-purity-energy (EPE)
%  phase diagram for 100000 randomly chosen two-qubit states where the entanglement is measured in terms of the concurrence. All quantities are dimensionless.} \label{figall}
%  \end{figure*}

\subsection{Separable States}

Separable states of a bipartite system may be expressed as a
convex sum of product states, \be \rho_{sep}=\sum_jp_j\rho_j^A
\otimes \rho_j^B\;\;.\ee This form is not particularly useful as
it is difficult in general to find such a decomposition. Here we
want to investigate the dependence of the purity and energy for
the separable states. In particular we would like to know the
bounds on these states i.e. given a separable state $\rho$ with
purity, $P$, what is the maximum energy $\rho$ can have? This is
equivalent to finding the minimum purity of separable state with a
certain purity. Given the Hamiltonian we use to calculate the
energy, it is reasonable to assume the separable states with
maximum or minimum energy for a given purity will be diagonal in
the basis $\{\ket{00},\ket{01},\ket{10},\ket{11}\}$.

% % Figure 3
% \begin{figure*}[floatfix]
%  \centering \setlength{\unitlength}{1cm}
%  %\includegraphics[width=8cm]{./paper_outline_neg.png}
%  \includegraphics[width=8cm]{./paper_outline_neg.eps}
%  \caption{(Color online) The boundaries of the entanglement-purity-energy (EPE) diagram for two-qubit states when the logarithmic negativity is used as the measure of entanglement. The MEMS
% are the Werner states from Eq. (\ref{werner}) in this case and form a line rather than a plane as in the case when concurrence is used to measure the entanglement. All quantities are dimensionless.} \label{outlineneg}
%  \end{figure*}

Thus we first consider the following density matrices, \be \rho=
\left(
\begin{array}{cccc} p_1&0&0&0\\ 0&p_2&0&0\\0&0&p_3&0\\0&0&0&p_1
\end{array}\right)\;\;, \ee with $E(\rho)=1,C(\rho)=0,P(\rho)=2p_1^2
+p_2^2+p_3^2$ and $2p_1+p_2+p_3=1$. States of this form include
the completely mixed state but the pure states $\ket{00}$ or
$\ket{11}$ cannot be expressed in this way. We shift the energy of
the state $\rho$ by a fixed amount, $\epsilon$, to get
$\rho_{\epsilon}$, \be \rho_{\epsilon}=\left(
\begin{array}{cccc} p_1+\epsilon&0 &0&0\\ 0&p_2-\epsilon&0&0 \\0&0&p_3&0
\\0&0&0&p_1
\end{array}\right)\;\;, \ee so that the energy is now $E(\rho_{\epsilon})=
1+\epsilon$ and the purity is $P(\rho_{\epsilon})=
(p_1+\epsilon)^2+ (p_2-\epsilon)^2 +p_3^2+p_1^2$. The minimum purity for a given energy, $E$, is thus $P_{min} =\frac{1}{4}(1+2(E-1)^2)$ for $\frac{1}{2}\le
E \le\frac{3}{2}$ to preserve positivity of the density
matrices. In order to calculate the minimum purity of states
$\rho$ with energy $E(\rho)\ge\frac{3}{2}$ we take the
following density matrices, \be \rho=\left(
\begin{array}{cccc} \epsilon&0 &0&0\\ 0&p-\epsilon&0&0 \\0&0&1-p&0
\\0&0&0&0
\end{array}\right)\;\;,\ee
where $\rho=\ket{00}\bra{00}$ when $\epsilon=p=1$. For these
states we have $E(\rho)=1+\epsilon$ and purity $P(\rho)=
(p-\epsilon)^2+(1-p)^2+\epsilon^2$. The minimum purity in this case is simply, \be P_{min}(\rho)=\frac{3}{2}E^2-2E+1\;\;,\ee with $0\le E\le\frac{1}{2}$.
Similarly, for $\frac{3}{2}\le E\le2$, 
%considering density matrices of the form, \be
%\rho=\left(
%\begin{array}{cccc} 0&0 &0&0\\ 0&p-\epsilon&0&0 \\0&0&1-p&0
%\\0&0&0&\epsilon
%\end{array}\right)\;\;,\ee
%where $\rho=\ket{11}\bra{11}$ when $\epsilon=p=1$, implies 
\be P_{min}(\rho)=\frac{3}{2}E^2 -4E+3\;\;.\ee
These three sections are
symmetric about $E=1$ as one would expect and $P_{min}=
\frac{3}{8}$ for each of the relevant expressions at
$E=\frac{1}{2}$ and $E=\frac{3}{2}$. This is the green
curve in Fig. 4 (curve satisfying $C=0$).
%curve in Fig. \ref{figoutline} (curve satisfying $C=0$).

% % Figure 4
% \begin{figure*}[floatfix]
% \centering
% %\setlength{\unitlength}{1cm}
% \subfigure[Boundaries of the set of physically allowed two-qubit states.]{\label{outline}
% %\includegraphics[width=4.5cm]{./paper_outline.png}}
% \includegraphics[width=4.5cm]{./paper_outline.eps}}
% \hspace{5mm}\subfigure[View of Fig. \ref{outline} looking along the Energy axis.]{ \label{CPout}
% %\includegraphics[width=4.5cm]{./paper_CP.png}}
% \includegraphics[width=4.5cm]{./paper_CP.eps}}
% \subfigure[View of Fig. \ref{outline} looking along the Purity axis.]{ \label{ECout}
% %\includegraphics[width=4.5cm]{./paper_EC.png}}
% \includegraphics[width=4.5cm]{./paper_EC.eps}}
% \hspace{5mm}\subfigure[View of Fig. \ref{outline} looking along the Concurrence axis.]{ \label{EPout}
% %\includegraphics[width=4.5cm]{./paper_EP.png}}
% \includegraphics[width=4.5cm]{./paper_EP.eps}}
% \caption{(Color online) The entanglement-purity-energy (EPE) phase diagram of the
% (i) Pure (blue curve) (ii) Separable (green curve) and
% (iii) maximally entangled mixed states (red curves) of two qubits. All quantities are dimensionless.}
% \label{figoutline}
% \end{figure*}

\subsection{MEMS}

The maximally entangled mixed states (MEMS) with respect to
concurrence and purity measures are conveniently expressed using
their concurrence, $C$, as a parameter, \bea
\non\rho_{MEMS}&=&C\ket{\phi_+}\bra{\phi_+}+
\frac{1}{3}\ket{01}\bra{01}\\&+&\non
\left(\frac{1}{3}-\frac{C}{2}\right)(\ket{00}\bra{00}+\ket{11}\bra{11})\;\;,
\eea for $C\in[0,\frac{2}{3}]$ while,
\be\rho_{MEMS}=C\ket{\phi_+}\bra{\phi_+}+(1-C)\ket{01}\bra{01}\;\;,\label{MEMS}\ee
for $C\in[\frac{2}{3},0]$, where $\ket{\phi_+}=
\frac{1}{\sqrt{2}}(\ket{00}+\ket{11})$. As they are written above
the MEMS have zero energy while the dependence of purity on the
concurrence is given by, \be P=\frac{1}{3}+\frac{C^2}{2}\;\;,\ee
for $C\in[0,\frac{2}{3}]$ and, \be P=C^2+(1-C)^2\;\;,\ee for
$C\in[\frac{2}{3},1]$. We can find the maximum energy these states
attain by applying local unitary operations, $U\in SU(2)\otimes
SU(2)$. Take $U=U_1(\theta_1,\vec{n}_1)\otimes
U_2(\theta_2,\vec{n}_2)$, with, \be
U_j=\exp\left[-i\theta_j(\vec{n}_j\cdot\vec{\sigma})\right]\;\;,\ee
where $|\vec{n}_j|=1$ for $j=1,2$ and
$\vec{\sigma}=(\sigma_x,\sigma_y,\sigma_z)$ are the Pauli
matrices. Applying $U$ gives $\rho'=U\rho_{MEMS}U^{\dagger}$ with
energy, \be E(\rho')=1+\frac{1}{3}\left[\sin^2\theta_2(1-n_{2,z}^2)-
\sin^2\theta_1(1-n_{1,z}^2)\right]\;\;.\ee Thus when
$C\le\frac{2}{3}$ the MEMS have an energy in the range
and $\frac{2}{3}\le E\le\frac{4}{3}$. For $C\ge \frac{2}{3}$, the same
calculation gives an energy for the locally transformed states,
\be E(\rho')=1+(1-C)\left[\sin^2\theta_2(1-n_{2,z}^2)-
\sin^2\theta_1(1-n_{1,z}^2)\right]\;\;,\ee and so $C\le E\le 2-C$ for these states. The MEMS
lines with maximum and minimum energy are the red curves in Fig. 4 (curves beginning at $E=1,P=1,C=1$).
%lines with maximum and minimum energy are the red curves in Fig. \ref{figoutline} (curves beginning at $E=1,P=1,C=1$).

%\subsection{EPE Diagram}
%\label{CPE2q}

The entanglement-purity-energy phase diagram is plotted in
Fig. 2 for 100,000 randomly generated density matrices
%Fig. \ref{figall} for 100,000 randomly generated density matrices
confirming the boundaries described above. There are some points on this diagram
worth highlighting, \begin{itemize} \item $\{C,P,E\}=\{1,1,1\}$
represents all the Bell states,
\bea\non\ket{\Phi^{\pm}}&=&\frac{1}{\sqrt{2}}
(\ket{00}\pm\ket{11})\;\;,\\\ket{\Psi^{\pm}}&=&\frac{1}{\sqrt{2}}
(\ket{01}\pm\ket{10})\;\;, \label{bell}\eea

\item $\{1,0,1\}$ represents all pure states of the form, \be
\ket{\psi} = a\ket{00}+b\ket{01}+c\ket{10}+a\ket{11}\;\;.\ee

\item $\{1,0,0\}$ represents $\ket{00}$.

\item $\{1,0,2\}$ represents $\ket{11}$.

\item $\{0,\frac{1}{4},0\}$ represents the completely mixed state,
$\rho=\frac{1}{4}\Id$.

\item $\{0,\frac{1}{4}\le P\le\frac{1}{3},1-\frac{1}{\sqrt{6}}\le
E\le 1+\frac{1}{\sqrt{6}}\}$ represents all the separable
states, including the completely mixed state, which cannot be
entangled by unitary operations.

\end{itemize}

\section{Two-mode Gaussian States}
\label{2modes}

We now consider a two-mode continuous variable system occupying a
Hilbert space which is the tensor product of two Fock spaces. The
creation and annihilation operators for the $j$th mode will be
denoted by $a_j^{\dagger}$ and $a_j$ respectively and these give
rise to the quadrature phase operators $x_j=(a_j+a_j^{\dagger})$
and $p_j=(a_j-a_j^{\dagger})/i$. We will only consider Gaussian
states of this system which are, by definition, states which have
a Gaussian characteristic or quasi-probability function in phase
space. For these states it is only necessary to specify the first
and second statistical moments of the quadrature operators.
Denoting $X=(x_1,p_1,x_2,p_2)$, to specify $\rho$ we need the mean
values $(\langle X\rangle)$ and the covariance matrix $\sigma$
where, \be \sigma_{ij}= \frac{1}{2}(\langle X_iX_j\rangle +\langle
X_jX_i\rangle)-\langle X_i\rangle\langle X_j\rangle \;\;. \ee Unitary
operations on $\rho$ correspond to symplectic operations on
$\sigma$ \cite{simukund}. Thus, \be U\rho
U^{\dagger}\Leftrightarrow S\sigma S^{\top}\;\;.\ee where $S\in
Sp(2,\mathbb(R))$ or $S\Omega S^{\top}=\Omega$ and, \be
\Omega=\left(\begin{array}{cc} \omega&0\\0&\omega\end{array}
\right)\;\;,\omega=\left (\begin{array}{cc} 0& 1\\-1&0\end{array}
\right)\;\;.\ee A local unitary operation $U_{loc}=U_1\otimes U_2$
correspond to the local symplectic operation $S_{loc}=S_1\oplus
S_2$ and each $S_j$ satisfies $S_j\omega S_j^{\top}$.

We can set the first moments of $X$ to zero by means of local
unitary operations which do not affect the entanglement or purity
properties of $\rho$. It has been shown that by local unitary
operations any CM $\sigma$ of a two-mode state can also be brought
to the so-called standard form \cite{simon,zoller}, \be
\sigma_{sf}=S^{\top}\sigma
S=\left( \begin{array}{cccc} a&0&c_+&0\\0&a&0&c_-\\ c_+&0&b&0\\
0&c_-&0&b\end{array}\right)\label{sf}\;\;.\ee

Writing $\sigma_{sf}\equiv\sigma$ (we drop the {\it sf} subscript
from now on for clarity) as \be \sigma_{sf}=S^{\top}\sigma
S=\left(
\begin{array}{cc}
\mathbf{\alpha}&\mathbf{\gamma}\\\mathbf{\gamma}^{\top}&
\mathbf{\beta}
\end{array}\right)\;\;,\ee the coefficients $a,b,c_+$
and $c_-$ are determined by the local symplectic invariants, \be
\mbox{Det }\alpha=a^2,\ \mbox{Det }\beta=b^2\ \mbox{and Det }
\gamma=c_+c_-\;\;.\ee The global symplectic invariants, invariant
under $Sp_{(4,\mathbb{R})}$ operations, are,
\begin{eqnarray}\nonumber \mbox{Det }\sigma&=&
(ab-c_+^2)(ab-c_-^2)\;\;,\\\Delta&=&\mbox{Det }\alpha+ \mbox{Det
}\beta+2\mbox{Det }\gamma\;\;,\label{seralian}\end{eqnarray} the latter being known in the literature as the {\it seralian} \cite{seralian} while
the former is related to the purity by, \be
P=\frac{1}{\sqrt{\mbox{Det }\sigma}}\;\;.\ee These local and
global invariants determine all the entanglement and mixedness
properties of a two-mode Gaussian state $\rho$.

Positivity of $\rho$ and the commutation relations for quadrature
phase operators mean that the CM $\sigma$ is required to satisfy
\cite{simukund,gaussind}, \be\sigma+i\Omega\ge 0\;\;,\label{HUR}\ee which in
turn means, \be\Delta\le1+\mbox{Det }\sigma\;\;,\ee in order to be
a proper covariance matrix. Any covariance matrix can be written
as \cite{sympleigs}, \be\sigma=S^{\top}\mathbf{\nu}S\;\;,\ee where
$S\in Sp_{(4,\mathbb{R})}$ and, \be\mathbf{\nu}=\mbox{diag}
(\nu_-,\nu_-,\nu_+,\nu_+)\;\;,\label{nu}\ee is the covariance matrix
corresponding to a tensor product of two thermal states. The state  $\mathbf{\nu}$ in Eq. (\ref{nu})
has a density matrix which is the tensor product of two thermal states,
\be
\rho=\rho_{th}^{(-)}\otimes\rho_{th}^{(+)}\;\;,\label{2Mth}\ee with, \be \rho_{th}^{(\pm)}
=\frac{2}{\nu_{\pm}+1}\sum_{j=0}^{\infty}
\left(\frac{\nu_{\pm}-1}{\nu_{\pm} +1}\right)^j\ket{j}\bra{j} \;\;,
\label{eq33}
\ee
and the average photon number in each of the thermal states, $\rho_{th}^{(\pm)}$, is
\be \bar{n}^{(\pm)}=\frac{\nu_{\pm}-1}{2}\;\;.\ee The quantities $\{\nu_+,\nu_-\}$ are
called the symplectic eigenvalues as they are invariant under
symplectic operations. In terms of the global invariants
$\mbox{Det}\sigma$ and $\Delta$ the symplectic eigenvalues have
the simple expression, \be 2\nu_{\pm}^2=\Delta\pm \sqrt{\Delta^2 -
4\mbox{Det}\sigma} \;\;.\ee Taking $\nu_-\le\nu_+$, Eq. (\ref{HUR}) reads, \be \nu_{-}\ge 1\;\;.\ee

We can now carry out a similar investigation of the relation
between the entanglement, purity and energy of two-mode quantum
states as we have shown for two qubits in the previous section.
This is the natural way to approach the concept of maximally
entangled states in continuous variable systems because, as
is well-known, any measure of entanglement can have arbitrarily large
values for a given purity. By restricting to a fixed energy we can
compare continuous
and discrete level systems on the same terms. First we define the
quantities under consideration in the case of two-mode Gaussian
states.

The quantification of entanglement of Gaussian bipartite states
has been studied extensively in recent years. One way has been to
utilize the partially transposed state (PPT) or Peres-Horodecki
condition for continuous variable systems \cite{simon}. The PPT
criterion is necessary and sufficient for the separability of two-mode Gaussian states 
and is defined as the transposition operation performed on
only one of the subsystems. In a two-mode CV system this amounts
to a mirror reflection in one of the
momenta operators, \be \rho\rightarrow\tilde{\rho}\Rightarrow
p_i\rightarrow -p_i \;\;,\ee for some $i=1,2$. In terms of local
the symplectic invariants this means that, \be\mbox{Det
}\gamma\rightarrow -\mbox{Det }\gamma\;\;. \ee And as this
operation does not preserve commutation relations, hence it is not
symplectic, the global invariants $\Delta$ and $\mbox{Det}\sigma$
change to, \bea\nonumber\tilde{\Delta}&=& \mbox{Det
}\alpha +\mbox{Det }\beta -2\mbox{Det }\gamma\\
\mbox{Det}\tilde{\sigma} &=& \mbox{Det} \sigma\;\;,\eea for the
new CM $\tilde{\sigma}$ corresponding to the transposed density
matrix $\tilde{\rho}$. The PPT criterion thus means that the
smallest symplectic eigenvalue of $\tilde{\sigma}$, \be
\tilde{\nu}_-= \sqrt{\frac{\tilde{\Delta}-\sqrt{\tilde{\Delta}^2-
4\mbox{Det}\sigma}}{2}}\;\;,\label{nuPPT}\ee must satisfy, \be
\tilde{\nu}_-\ge 1\;\;.\label{entnumin}\ee In fact, this
eigenvalue completely quantifies the entanglement through its
relation with the negativity,
\be\mathcal{N}=\max\left[0,\frac{1-\tilde{\nu}_-}
{2\tilde{\nu}_-}\right]\;\;,\ee introduced in Ref. \cite{neg} and
applied to CV systems in Ref. \cite{negcv}, and the logarithmic
negativity, \be
\mathcal{E}_{\mathcal{N}}(\rho)=\max\{0,-\ln\tilde{\nu}_-\}\;\;,\ee
which are both decreasing functions of $\tilde{\nu}_-$, thus
quantifying the violation of Eq. (\ref{entnumin}).

Recently, Wolf {\it et al.} \cite{wolfEOF} have defined the
entanglement of formation (EoF) for Gaussian states and, further,
have shown how to explicitly calculate this quantity for a given
CM $\sigma$. It is an entanglement monotone under local Gaussian operations and classical communication. This quantity has been further studied in Ref.
\cite{adess} and there it is shown that the entanglement of
formation leads to a different ordering of the states compared to
that induced by the negativity above. We could use the
entanglement of formation here but, apart from the symmetric two-mode Gaussian states, there is no analytical
expression for general Gaussian states. 
%However, for symmetric
%two-mode Gaussian states the EoF is given in terms of the smallest
%symplectic eigenvalue of the transposed CM $\tilde{\sigma}$.
% by,
%\be E_F= \max[0,h(\tilde{\nu}_-)]\;\;,\ee where, \be h(x)=
%\frac{(1+x)^2}{4x} \ln \left[\frac{(1+x)^2}{4x} \right]
%-\frac{(1-x)^2}{4x}
%\ln\left[\frac{(1-x)^2}{4x}\right]\;\;.\label{eqnh}\ee 
In fact for arbitrary two-mode Gaussian states, finding the EoF is quite a non-trivial
task. The solution has a nice geometric description
which is discussed in Refs. \cite{wolfEOF,adess}. Ultimately,
though, it is determined via numerical minimisation so we choose
to deal with the logarithmic negativity due to its more analytical
nature.

To characterize the mixedness we use the purity again which for
two-mode pure states is simply given by, \be
P(\sigma)=\frac{1}{\sqrt{\mbox{Det }\sigma}}\;\;.\ee

For the energy we will use the sum of the average number of
photons present in each mode, \be E(\rho)=\langle
a_1^{\dagger}a_1\rangle + \langle a_2^{\dagger}a_2\rangle=
\bar{n}_1+\bar{n}_2\;\;.\ee In terms of the CM, $\sigma$,
corresponding to the state $\rho$, \be
E(\rho)=\frac{1}{4}\mbox{Tr}\sigma -1+\sum_j\langle X_j\rangle^2\;\;.\ee As
we have mentioned, the mean values, $\langle X_j\rangle$, can be set to
zero by local operations which do not affect the entanglement or purity.
These operations obviously change the energy though, so while we set them to
zero in the following, the energies calculated are the minimum values for a
given $\sigma$ in the standard form. As an example all two-mode coherent states
are represented by the vacuum, with zero energy.

% % Figure 5
% \begin{figure*}[floatfix]
% \centering \setlength{\unitlength}{1cm}
% %\includegraphics[width=8cm]{./paper_2M_entE.png}
% \includegraphics[width=8cm]{./paper_2M_entE.eps}
% \caption{Graph depicting how the energy, $E$, of pure two-mode
% Gaussian states depends on the amount entanglement, $\mathcal{E}$, between the two
% modes. The physical region lies in the shaded region above the curve. All quantities are dimensionless.} \label{figpure2M}
% \end{figure*}

The local invariants $a$ and $b$ of a CM $\sigma$ are related to
these expectation values of the number operators of the two modes
by \be a=2\bar{n}_1+1\;\;, \;\;b=2\bar{n}_2+1\;\;.\ee Also $a$ and
$b$ are the inverses of the purities of the reduced density
matrices of the two modes, $a =1/\mu_1=1/\mbox{Tr} \rho_1^2$,
$b=1/\mu_2=1/\mbox{Tr} \rho_2^2$ with $\rho_{1,2}= \mbox{Tr}_{2,1}
\rho$. This fact has been used to characterize the maximally and
minimally entangled two-mode mixed Gaussian states, for fixed
global and local purities \cite{seradill}, where the entanglement
is quantified using the negativity.
%There is another parametrization introduced
%in \cite{adess} where $a$ and $b$ are given by $a=s+d$ and
%$b=s-d$. These parameters have a physical interpretation as the
%sum of the average number of photons in each mode (plus 1) and the
%difference in the these average photon numbers, \be
%s=\bar{n}_1+\bar{n}_2+1\;\;,\;\;d=\bar{n}_1-\bar{n}_2\;\;,\ee an
%interpretation strongly hinted at in the choice of variable names
%although not mentioned in the paper.

We will now outline the EPE diagram for the two-mode Gaussian
states by first examining the extremal states. These states have
been studied in depth in Refs. \cite{seradill,adess}. In Ref. \cite{seradill}
the authors study the extremal entanglement and mixedness of
two-mode Gaussian states. They introduce generalized maximally and
minimally entangled mixed states, denoted GMEMS and GLEMS
respectively, which are maximally and minimally entangled states
for fixed global and local purities. Given how the local purities
are defined, this would appear to be equivalent to finding the
maximally and minimally entangled states for a fixed purity and
average number of photons in each mode. However, we will see that
the minimally entangled mixed states for fixed energy are simply
the separable states.

In the following we will restrict ourselves to the range of energies,
$0\le E\le 2$, where the energy is measured in dimensionless units (number of excitations),
in order to draw parallels with the two-qubit case.

\subsection{Pure States}

A pure, $P=1$, two-mode Gaussian state has a symmetric CM,
corresponding to a two-mode squeezed vacuum, with $a=b$ and
satisfies $c_+=-c_-=\sqrt{a^2-1}$. The minimum energy is simply, \be
E_{min}(\rho)=a-1\;\;.\ee There is thus a simple form for the symplectic
eigenvalue of the partially transposed density matrix, \be
\tilde{\nu}_- =a-\sqrt{a^2-1}\;\;.\ee The dependence of the logarithmic
negativity for pure states on the energy is shown in Fig. 5 up to a maximum energy of 2. Any point on the curve in Fig. 5 represents a pure state with CM
%negativity for pure states on the energy is shown in Fig. \ref{figpure2M} up to a maximum energy of 2. Any point on the curve in Fig. \ref{figpure2M} represents a pure state with CM
given by the above values. However by applying a local squeezing
operator of the form, \be
S_{sq}=\left(\begin{array}{cccc}e^{r}&0&0&0\\0&e^{-r}&0&0\\0&0&
e^{-r}&0\\0&0&0&e^{r}\end{array}\right)\;\;,\label{squeeze}\ee we
can change the energy of the state to, \be
E(U(S_{sq})^{\dagger}\rho U(S_{sq}))=a\cosh 2r-1\;\;,\ee so that
the curve shown indicates the maximally entangled pure two-mode
states for the given energy, i.e. the two-mode squeezed vacua,
and all pure states above the curve are physically allowed.

\subsection{Separable States}

The separable states are all those for which the smallest
symplectic eigenvalue of the partially transposed CM satisfies,
\be \tilde{\nu}_-\ge 1\;\;,\ee or in other words the partially
transposed covariance matrix is still a valid one. The separable
two-mode Gaussian states are determined by the value of the energy
through the condition, \be P\ge\frac{1}{ab}\;\;,\ee which is a
direct consequence of keeping $\sigma\ge 0$, and implies that \be
P\ge \frac{1}{(E+1)^2}\;\;,\label{sepP}\ee for separable states.
The inequality (\ref{sepP}) comes from considering all the
combinations of ``local'' energies in each mode which added
together give $E$. The minimum occurs when both modes contribute
equally i.e. for symmetric states. These states are the tensor product of two
thermal states in Eq. (\ref{2Mth}) with equal average photon number in both modes.
The curve shown in Fig. 6 shows the minimum energy a separable two-mode Gaussian state can have
%The curve shown in Fig. \ref{figsep2M} shows the minimum energy a separable two-mode Gaussian state can have 
for a given purity and thus, as above, all the physical states lie above this curve.

% % Figure 6
% \begin{figure*}[floatfix]
% \centering \setlength{\unitlength}{1cm}
% %\includegraphics[width=8cm]{./paper_2M_EP.png}
% \includegraphics[width=8cm]{./paper_2M_EP.eps}
% \caption{Graph showing how the energy, $E$, of the separable
% two-mode Gaussian states depends on the purity, $P$, for energy values
% up to 2. The physical region lies in the shaded region above the curve. All quantities are dimensionless.} \label{figsep2M}
% \end{figure*}

\subsection{GMEMS \& GLEMS}

The Gaussian maximally and minimally entangled states for fixed global and
local purities were
introduced by G. Adesso {\it et al.} in Refs. \cite{seradill1,seradill}
through a parametrization of two-mode Gaussian states in terms of
the local purities, \be \mu_1=\frac{1}{a},\;\; \mu_2= \frac{1}{b}
\;\;, \ee and the global symplectic invariants, the purity and
seralian. By writing the parameters $a$, $b$ and $c_{\pm}$ in
terms of the above parameters, the seralian must satisfy, \be
\Delta_{min}\le\Delta\le \Delta_{max}\;\;,\label{seralminmax}\ee
where $\Delta_{min}$ and $\Delta_{max}$ are easily calculated.
From Eq. (\ref{nuPPT}) it is seen that, \be
\left.\frac{\partial \tilde{\nu}_-^2}{\partial
\Delta}\right|_{\mu_1, \mu_2,P}>0\;\;,\ee so that the entanglement
of a state as measured by the negativity or logarithmic negativity
is a monotonically decreasing function of the seralian. Thus the
limits imposed by Eq. (\ref{seralminmax}) imply that there exist
maximally and, surprisingly, minimally entangled two-mode Gaussian
states for fixed values of the global and local purities.

Here we are looking for the maximally entangled mixed Gaussian two-mode states for fixed energy.
They are found by again considering a state, $\rho$, with CM $\sigma$ in the standard form (\ref{sf}).
We now fix the energy of the state, \be E=\frac{1}{2}(a+b)-1\;\;,\label{Ee}\ee
and the purity, \be
P=\frac{1}{\sqrt{(ab-c_+^2)(ab-c_-^2)}}\;\;.\label{Pe}
\ee It is easily seen from Eq. (\ref{nuPPT}), and pointed out in Ref. \cite{seradill1}, that the smallest
symplectic eigenvalue of the CM
of the partially transposed state, $\tilde{\nu}_-$, satisfies, \be
\left.\frac{\partial \tilde{\nu}_-^2}{\partial \tilde{\Delta}}\right|_{E,P}<0\;\;,\ee
so that the entanglement of a given state is
monotonically increasing with the seralian of the CM corresponding to the
partially transposed state. Thus, in order to find the states with
maximum entanglement for a given purity and energy, we must find where
$\tilde{\Delta}$ has a maximum. Manipulating Eqs. (\ref{Ee}) and (\ref{Pe})
gives, \be \tilde{\Delta}=4(E+1)^2-2\sqrt{\frac{1}{P^2}+ab(c_++c_-)^2}
\;\;,\ee which is maximised when $c_+=-c_-$ and thus the maximally entangled
mixed Gaussian states satisfy, \be\tilde{\Delta}=4(E+1)^2-\frac{2}{P}\;\;
\label{Gmem}.\ee These are shown in Fig. 7 for energy 2 (in dimensionless units).
%\label{Gmem}.\ee These are shown in Fig. \ref{figgmems} for energy 2 (in dimensionless units).

% % Figure 7
% \begin{figure*}[floatfix]
% \centering \setlength{\unitlength}{1cm}
% %\includegraphics[width=8cm]{./paper_2M_entP.png}
% \includegraphics[width=8cm]{./paper_2M_entP.eps}
% \caption{Graph showing how the entanglement, $\mathcal{E}$, of two-mode
% Gaussian states depends on the purity, $P$, for a fixed energy value of
% 2. The upper curve indicates the maximally entangled mixed states for fixed energy and the physical region is the shaded area below this curve. All quantities are dimensionless.} \label{figgmems}
% \end{figure*}

The Gaussian maximally entangled mixed states for fixed global and
local purities, or GMEMS, as defined in Refs. \cite{seradill1,seradill}
have a simple expression for states with a standard form CM. These
states saturate the lower bound in (\ref{seralminmax}) and this
means that, \be c_{\pm}=\pm\sqrt{ab-\frac{1}{P}}\;\;,\ee and
importantly for the definition of entanglement,
\be\tilde{\Delta}=4(E+1)^2-\frac{2}{P}\;\;,\label{tildedel}\ee in agreement
with Eq. (\ref{Gmem}).
Thus, not unexpectedly, the maximally entangled states for a given purity and
energy coincide with the GMEMS.

In Fig. 7 one notices the fact that there is no
%In Fig. \ref{figgmems} one notices the fact that there is no
entanglement below a certain purity value. This arises from the
condition that $\tilde{\nu}_-\ge1$ which, together with Eq. (\ref{Gmem}),
means that, \be P\ge\frac{1}{2E+1}\;\;,\ee in order to be able to unitarily generate
entangled states. This bound and the condition in Eq.
(\ref{sepP}) defines a range of values of the purity of two-mode
Gaussian states, \be \frac{1}{(E+1)^2}\le
P<\frac{1}{2E+1}\;\;,\label{Pbound}\ee such that, for a given
energy $E$, there can be no entanglement in these states. This is
similar to the case of two qubits where if the purity is lower than
one third, only separable states exist independent of the energy.
Physically, it means that in this range for a given value of
energy, the states are simply too classically mixed to accommodate
quantum correlations. The range of values for the purity increases
from zero, at $E=0$, reaches a maximum when the energy attains the
golden ratio, \be E=\frac{1+ \sqrt{5}}{2} \;\;,\ee and then falls
to zero again as the energy increases to infinity.

It is an interesting coincidence that for the case of $E=1$ both the 2-mode
and 2-qubit upper and lower bounds agree, $\frac{1}{4}\le P
\le\frac{1}{3}$. This suggests the maximally
mixed two-mode Gaussian state, of a given energy $E$, is that
state which saturates the lower bound in Eq. (\ref{Pbound}). Such a
state would have a CM, \be\sigma_{mixed}=(E+1)\mathbb{I}_4\;\;,\label{2Mmix}\ee where
$\mathbb{I}_4$ is the $4\times4$ identity matrix, in analogy with
the two-qubit case, and corresponds to a state which is a tensor
product of two thermal states with equal average number of photons
in each mode as we have seen before in Eq. (\ref{2Mth}).

The continuous variable version of the Werner states of two qubits in Eq. (\ref{werner}) were introduced in Ref. \cite{CVwerner} and have the form, \be \rho_{GW}=
r\rho_{ME}+(1-r)\rho_I\;\;,\ee where $\rho_{ME}$ is the pure two-mode Gaussian state with
energy $E$ and $\rho_I$ is the completely mixed state specified by the CM in Eq. (\ref{2Mmix}),
i.e. the density operator $\rho_I$ is equal to a tensor product of two thermal density operators
each corresponding to a state with the average photon number equal to $\bar{n}=E/2$ (we remind us that that we consider units such that $\hbar=\omega=1$),
i.e.,
\be \rho_{GW}=
r\rho_{ME}+(1-r)\rho_{th}\otimes\rho_{th}\;\;,
\ee
where $\rho_{th}$ is given by Eq.~(\ref{eq33}) in terms of symplectic eigenvalues, or in terms of the mean excitation number $\bar{n}$ we have
\be
\rho_{th}= \frac{1}{(\bar{n}+1)}\sum_{j=0}^\infty \left(\frac{\bar{n}}{\bar{n}+1}\right)^j |j\rangle\langle j|
\; \; . \ee These states are not Gaussian however and thus do not appear on our EPE phase diagram.

% % Figure 8
% \begin{figure*}[floatfix] \centering
% \setlength{\unitlength}{1cm}
% %\includegraphics[width=8cm]{./paper_2M_entP_GLEMS.png}
% \includegraphics[width=8cm]{./paper_2M_entP_GLEMS.eps}
% \caption{Graph showing the GMEMS and GLEMS of two-mode Gaussian
% states for fixed energy value of 2. The GMEMS are those states lying on the upper curve while the GLEMS are those states which lie in the darker shaded region between the two curves. All quantities are dimensionless.} \label{figgmgl}
% \end{figure*}

The GLEMS are states which by definition saturate the upper bound
in Eq. (\ref{seralminmax}). From Refs. \cite{seradill1,seradill} states
satisfy this condition if and only if the symplectic spectrum is
$\nu_-=1,\nu_+=1/P$ and this leads to, \be
\mbox{Det}\gamma=\frac{1}{ab}\left[\left(\Delta-(a-b)^2 \right)^2
- \left(\Delta-(a+b)^2\right)^2 \right]\;\;,\ee for these states
described by a standard form CM and thus,
\be\tilde{\Delta}=4(E+1)^2-\left(1+\frac{1}{P^2}\right)
\label{glemsdel}\;\;.\ee The first term arises out of the fact
that the minimum value of $\tilde{\Delta}$ occurs when the states
are described by a symmetric CM. The GLEMS actually describe a
range of states of which those with the minimum entanglement are
shown in Fig. 8 for energy 2 and they exist in the darker shaded region between the two curves shown. The GLEMS and GMEMS coincide when the global
%shown in Fig. \ref{figgmgl} for energy 2 and they exist in the darker shaded region between the two curves shown. The GLEMS and GMEMS coincide when the global
purity can be written in terms of the two local purities and these
states are called GMEMMS or Gaussian maximally entangled mixed
states for fixed marginals. As noted in Refs. \cite{seradill1,seradill},
the GLEMS contain separable states which are readily seen in Fig. 8. The range of purities for which these
%the GLEMS contain separable states which are readily seen in Fig. \ref{figgmgl}. The range of purities for which these
states can tolerate separable states is, for a given energy $E$, \be \frac{1}{2E+1}\le P\le
\frac{1}{\sqrt{2E^2 +4E+1}}\;\;.\ee

We can now put together the pieces of the EPE diagram for 2-mode
Gaussian states. The resulting graph, shown in Fig. 9, follows the outline of a quarter of a fruit bowl.
%Gaussian states. The resulting graph, shown in Fig. \ref{figCPE2M}, follows the outline of a quarter of a fruit bowl.
Now in this picture the idea of minimally entangled mixed states
for a fixed energy has no meaning, unless you consider the
separable states as such. Thus the GLEMS do not arise naturally in our picture.
This is understandable as we are using only three
parameters to describe the set of states instead of four and thereby lose some information.

%The GMEMS and GLEMS can be written in their standard form with,
%$a=E+d+1$, $b=E-d+1$ and, \begin{eqnarray}\nonumber
%c_{\pm}&=&\frac{1}{4\sqrt{(E-1)^2-d^2}}\left(\sqrt{(4d^2-\Delta)^2-\frac{4}{P^2}}\right.\\
%&&\left.\pm\sqrt{(4(E-1)^2
%-\Delta)^2-\frac{4}{P^2}}\right)\;\;,\end{eqnarray}
%with, \be \Delta=\left\{\begin{array}{c}2\left(\frac{1}{P}+2d^2\right)\;\;,\mbox{ for GMEMS}\;\;.\\
%\left(\frac{1}{P^2}+1\right)\;\;,\mbox{ for
%GLEMS}\;\;.\end{array}\right.\ee

% \section{EPE Diagram: 2-mode Gaussian States}
%\label{CPE2m}

% Figure 9
% \begin{figure*}[floatfix]
% \centering \setlength{\unitlength}{1cm}
% %\includegraphics[width=8cm]{./paper_2M_entPE.png}
% \includegraphics[width=8cm]{./paper_2M_entPE.eps}
% \caption{(Color online) The entanglement-purity-energy (EPE) phase diagram boundaries showing the (i) pure (blue curve satisfying $P=1$) (ii) separable (green curve satisfying $\mathcal{E}=0$) and (iii) maximally entangled mixed (red curves starting on the pure curve) Gaussian states of a two-mode continuous variable
% system for energies, $E$, in the range $0\le E\le 2$. All quantities are dimensionless.} \label{figCPE2M}
% \end{figure*}

Some points on this diagram worth highlighting are the following, \begin{itemize} \item
$\{C,P,E\}=\{0,\frac{1}{(E+1)^2},E\}$ represents all the
completely mixed two-mode Gaussian state.

\item $\{0,1,0\}$ represents all tensor product of the vacuum
state of two modes

\item $\{0,\frac{1}{(E+1)^2}\le P\le\frac{1}{2E+1},E\}$ represents
all the separable states, including the completely mixed state,
which cannot be entangled by arbitrary unitary operations.

\end{itemize}

\section{Entanglement Transfer: 2 modes to 2 qubits}
\label{CPE2q2m}
Now we complete our study of the two sets of states, those of two qubits compared to the two-mode Gaussian states, by considering how efficiently one may transfer
entanglement and energy from a two-mode field to two qubits. In Refs. \cite{maur1,maur2,maur3,maur4} entanglement swapping between qubits and continuous variables has been investigated in detail. Here we are particularily interested in the role of energy in the entanglement transfer process.
We will employ the basic resonant Jaynes-Cummings Hamiltonian \cite{JC}, \be H_{JC}=\sum_j\lambda_j(a_j\sigma_+^{(j)}+a^{\dagger}_j\sigma_-^{(j)})\;\;,\label{JC}\ee
where $a^{\dagger}_j$ and $a_j$ are the creation and annihilation operators for the $j$th mode and  $\sigma_{\pm}^{(j)}$ are the raising and lowering operators for the $j$th atom, %$\sigma_{-}\ket{e}=\ket{g},\sigma_{+}\ket{g}=\ket{e}$,
to describe the interaction of the two
modes with the two two-level atoms. The situation envisaged is that the first mode interacts with the first atom only and similarily
for the second mode and the second atom with both atom-field couplings equal, $\lambda_1=\lambda_2=\lambda$. We will analyse three different
initial states of the field while the state of the two atoms will always be in the ground state ($\ket{g}_1\ket{g}_2$) initially.

First we consider when the field is in a superposition of one photon in the first mode and zero photons in the second and vice versa,
\be\ket{\psi_{f}(t=0)}=\frac{1}{\sqrt{2}}(\ket{0}_1\ket{1}_2+\ket{1}_1\ket{0}_2\;\;.\ee 
There has been some recent discussion
\cite{pawcza,vanenk} related to this state. Intuitively this state is an entangled state of two modes and should generate entanglement between the two qubits. Indeed, as the input state is pure, in this case and all the following cases, we can use the von Neumann entropy of the reduced state of one of the modes as a measure of the entanglement of this state, \be S(\ket{\psi_{f}(0)})=-{\mbox Tr}\rho_j\ln\rho_j,\;\;j=1,2\;\;,\ee where $\rho_j$ is the reduced density matrix of the first or second mode.

The initial energy in the two-mode field is one excitation and the entropy is $S(\ket{\psi_{f}(0)})=\ln 2$. When we allow the field and atoms interact for a time, $t$, under the Hamiltonian in Eq. (\ref{JC}), the combined output state for the two-mode field and the two qubits is, \bea\nonumber\ket{\psi_{f,at}(t)}&=&\cos(\lambda t)\ket{\psi_{f}(0)}\ket{gg}\\&-i&\sin(\lambda t)\ket{0}_1\ket{0}_2
\frac{(\ket{ge}+\ket{eg})}{\sqrt{2}}\;\;,\eea where $\ket{0}_j$ represents the vacuum state of the $j$th mode. Thus at any time $t$ such that $gt=\frac{n\pi}{2}$ the two qubits are in a maximally entangled Bell state. At these times the initial energy in the field is completely transferred to the state of the atoms and the joint state of the atoms and two-mode field is separable and pure. What about a similar but more general input field state of the form, \be
\ket{\psi_{f}(0)}=\frac{1}{\sqrt{2}}(\ket{0}_1\ket{n}_2+\ket{n}_1\ket{0}_2?\;\;\ee
The initial entanglement is $\ln 2$ as previously but there are now $n$ excitations present. It turns out that, after the same evolution under Hamiltonian in Eq. (\ref{JC}), the resulting two-qubit density matrix is given by, \be \rho(t)=\frac{1}{2}\left(\begin{array}{cccc}
2\cos^2\lambda t\sqrt{n}&0&0&0\\0&\sin^2\lambda t\sqrt{n}&\delta_{n,1}\sin^2\lambda t\sqrt{n}&0\\0&\delta_{n,1}\sin^2\lambda t\sqrt{n}&\sin^2\lambda t\sqrt{n}&0\\0&0&0&0\end{array}\right)\;\;,\ee so that for $n>1$ no entanglement is transferred from the two-mode field to the pair of atoms and their state remains separable even though energy is continuously swapped between the field and the atoms.
% % Figure 10
% \begin{figure*}[floatfix] \centering
% \setlength{\unitlength}{1cm}
% %\includegraphics[width=8cm]{./paper_2M_Ein_entout1.png}
% \includegraphics[width=8cm]{./paper_2M_Ein_entout1.eps}
% \caption{The dependence of the maximum concurrence, $C_{out}$, generated between two two-level atoms by evolution under the Jaynes-Cummings Hamiltonian (\ref{JC}), on the initial energy, $E_{in}$, in the two mode field. The field is initially in the entangled superposition state of Eq. (\ref{entcoh}) characterised by the coherent state amplitude $\alpha$. All quantities are dimensionless.} \label{figEinentout}
% \end{figure*}

The next case we consider is the input field,
\be\ket{\psi_{f}(0)}=\frac{1}{\sqrt{2}\sqrt{1+e^{-|\alpha|^2}}}(\ket{0}_1\ket{\alpha}_2+
\ket{\alpha}_1\ket{0}_2)\;\;,\label{entcoh} \ee where $\ket{\alpha}$ is a coherent state. The initial energy in this state, $\bar{N}=\langle N_1+N_2\rangle$, is, \be\bar{N}=\frac{|\alpha|^2}{1+e^{-|\alpha|^2}}\;\;,\ee and the initial entanglement ranges from 0 to $\ln 2$ depending on the value of $\alpha$. After evolution under Eq. (\ref{JC}) for a time $t$, the state of the two qubits is well-described by a density matrix of the form, \be\rho(t)=\left(
\begin{array}{cccc}a&x&y&0\\x^*&b&z&0\\y^*&z&b&0\\0&0&0&0\end{array}\right)\;\;\ee with
\be z=\frac{|\alpha|^2e^{-|\alpha|^2}}{2(1+e^{-|\alpha|^2})}\sin^2(\lambda t)\;\;.\ee The concurrence of
this state is $C=2z$ which is maximised when there is on average one excitation initially in the field modes, i.e. $\bar{N}=1$. This is shown in Fig. 10 where the  maximum entanglement generated 
%field modes, i.e. $\bar{N}=1$. This is shown in Fig. \ref{figEinentout} where the  maximum entanglement generated 
between the two atoms is plotted as a function of the initial energy in the field. We have also plotted in Fig. 11 the dependence of the maximum entanglement generated between the two atoms on 
%plotted in Fig. \ref{figentinentout} the dependence of the maximum entanglement generated between the two atoms on 
the initial entanglement present between the two modes. It is interesting that for the initial state of the two modes in Eq. (\ref{entcoh}) the degree of entanglement of this state need not be the maximum allowed in order to achieve the maximum amount of entanglement possible between the two atoms. The fact that the entanglement between the two atoms is maximised when there is on average one excitation initially in the field is not so surprising given the previous input states considered. In any case, the amount of entanglement generated between the two atoms is very small.
% % Figure 11
% \begin{figure*}[floatfix] \centering
% \setlength{\unitlength}{1cm}
% %\includegraphics[width=8cm]{./paper_2M_entin_entout1.png}
% \includegraphics[width=8cm]{./paper_2M_entin_entout1.eps}
% \caption{The dependence of the maximum concurrence, $C_{max}$, generated between two two-level atoms by evolution under the Jaynes-Cummings Hamiltonian (\ref{JC}), on the initial entanglement in the two mode field. The field is initially in the entangled superposition state of eqn. (\ref{entcoh}) characterised by the coherent state amplitude $\alpha$. All quantities are dimensionless.} \label{figentinentout}
% \end{figure*}

Finally we consider the case when the input state of the field is a pure two-mode Gaussian state,
\be \ket{\psi_{f}(0)}=\sqrt{1-\gamma^2}\sum_n\gamma^n\ket{n}_1\ket{n}_2\;\;,\ee with $\gamma=\tanh r$ where the above state can be generated by applying the two-mode squeezing operator, $S(r)=\exp[-r(a^{\dagger}_1a^{\dagger}_2-a_1a_2)]$, to the two-mode vacuum state. The initial energy in the field is given by, \be E_{in}=\frac{2\gamma^2}{1-\gamma^2}\;\;,\ee and the initial entanglement is, \be S_{in}=\ln(1-\gamma^2)+E_{in}\ln\gamma\;\;.\ee The state of the two qubits after a time evolution is given by, \be\rho(t)=\left(
\begin{array}{cccc}a&0&0&x\\0&b&0&0\\0&0&b&0\\x&0&0&d\end{array}\right)\;\;,\ee where, \bea
\nonumber a&=&\frac{1}{4}A\sum_n\gamma^{2n}(\cos2\sqrt{n}\lambda t+1)^2\;\;,\\
\nonumber b&=&\frac{1}{4}A\sum_n\gamma^{2n}\sin^22\sqrt{n}\lambda t\;\;,\\
\nonumber d&=&\frac{1}{4}A\sum_n\gamma^{2n}(\cos2\sqrt{n}\lambda t-1)^2\;\;,\\
\nonumber x&=&\frac{1}{4}A\sum_n\gamma^{2n+1}(\cos2\sqrt{n+1}\lambda t-1)\\
&&\hspace{20mm}\times(\cos2\sqrt{n}\lambda t+1)\;\;,\eea and
$A=(1-\gamma^2)$. In Fig. 12 the maximum entanglement, $C_{out}$, generated between the pair of 
%$A=(1-\gamma^2)$. In Fig. \ref{figEinCPout} the maximum entanglement, $C_{out}$, generated between the pair of 
atoms and the purity, $P$, of the state of the two atoms at this time is plotted as a function of the initial energy in the field, $E_{in}$. The maximum entanglement generated between the atoms is quite high, $C_{out}^{max}\sim 0.9$, in this case but it does not occur when there is initially one excitation in the field as was the case for the previous input states. In Fig. 13 we plot the dependence of the maximum entanglement generated on the initial entanglement in the field. Again we see that it is not necessary to initially have the most entangled two-mode state possible in order to generate the most maximally entangled two-qubit state possible by evolution under the resonant Jaynes-Cummings Hamiltonian.

% % Figure 12
% \begin{figure*}[floatfix] \centering
% \setlength{\unitlength}{1cm}
% %\includegraphics[width=8cm]{./paper_2M_Ein_CPout.png}
% \includegraphics[width=8cm]{./paper_2M_Ein_CPout.eps}
% \caption{(Color online) The dependence of the maximum concurrence, $C_{out}$ (lower curve), generated in a pair of two-level atoms by evolution under the Jaynes-Cummings Hamiltonian (\ref{JC}), and the purity, $P$ (upper curve), of the state of the two atoms at this time on the initial energy, $E_{in}$, of the two-mode field when the field is initially in a pure two-mode Gaussian state. All quantities are dimensionless.} \label{figEinCPout}
% \end{figure*}

% % Figure 13
% \begin{figure*}[floatfix] \centering
% \setlength{\unitlength}{1cm}
% %\includegraphics[width=8cm]{./paper_pure2M_Sin_Cout.png}
% \includegraphics[width=8cm]{./paper_pure2M_Sin_Cout.eps}
% \caption{The dependence of the maximum concurrence, $C_{out}$, generated in a two-qubit pair by evolution under the Jaynes-Cummings Hamiltonian (\ref{JC}), on the initial entanglement in the two-mode field, $S_{in}$, when the field is initially in a pure two-mode Gaussian state. All quantities are dimensionless.} \label{figSinCout}
% \end{figure*}

\section{Conclusion}
\label{conc}

We have set out to show that the energy of a state is an important
consideration when specifying the maximally entangled mixed
states of either two qubits or a two-mode continuous variable system.

For two qubits there is a continuous range of MEMS lines on the entanglement-purity-energy (EPE)
diagram, each depending on the average number of excitations in
the two-qubit system. These states, together with the separable
and pure states, form the surface of the volume containing all
physically allowed quantum states. The states in (\ref{MEMS}) are
those which maximise the concurrence for a given purity {\it and}
have exactly one excitation. They include the Bell states
(\ref{bell}). If a state has $E\neq 1$ then it can still be considered a
maximally entangled state albeit one less entangled than the
states (\ref{MEMS}) for the same purity.

For the case of two-mode states it is natural that the energy be fixed in order to
specify the maxmally entangled mixed states. Through this parametrization we have provided a nice way view the set of two-mode Gaussian states. From this EPE diagram the set of physically allowed states of these two fundamentally different systems can be compared. There are some interesting similarities we have pointed out which go beyond the fact that both systems are completely specified by a $4\times4$ matrix. The EPE diagrams are also an aide to view the allowed dynamical evolution of these states.

Finally we have looked at how entanglement can be swapped between these two systems using
a simple resonant Jaynes-Cummings model. We have shown that the initial energy and the initial entanglement present in the two-mode field are important factors as to how efficiently the entanglement is transferred from one system to the other. Physically this is a method of distributing entanglement between two distant atomic systems and what is interesting is that the two-mode Gaussian states are very effective at entangling the two qubits. Given that these states are experimentally less demanding to produce, this is a nice result. The natural extension of this idea to a two-mode field interacting with $N$ atoms may serve as a quantum memory \cite{EPJF} and this will be studied later \cite{DMMZVB}.

{\bf Acknowledgement}\\
 This work was supported in part by the
European Union projects INTAS-04-77-7289, CONQUEST and QAP,  by
the Slovak Academy of Sciences via the project CE-PI I/2/2005, by the
project APVT-99-012304. One of us (V.B.) acknowledges the support from the Alexander von Humboldt
Foundation.

%\begin{table}[H]
%\begin{center}
%\begin{tabular}{|l|c||l|c|}
%\hline
%3-pulse&C&5-pulse&C\\
%\hline
%$W_1^{\epsilon}$(BB1)&4.7&$W_{121}^{\epsilon}$&72.3\\
%\hline
%$W_2^{\epsilon}$(PB1)&59.1&$W_{112}^{\epsilon}$ &190.6\\
%\hline
%$W_3^{\epsilon}$&283.4&$W_{222}^{\epsilon}$ & 877.8\\
%\hline
%\end{tabular}
%\caption{The coefficients $C$ in the fidelity expansion $F=1-C\epsilon^6$
%for six composite pulse sequences which compensate for an
%error-prone $\pi$-pulse around the -$X$-axis.}
%\label{table1}
%\end{center}
%\end{table}

%\begin{figure}[p]
%\begin{center}
%\setlength{\unitlength}{1cm}
%\begin{picture}(6,10)
%\put(-2.5,.5){\includegraphics[width=10cm]{fidfelity.eps}}
%\end{picture}
%\end{center}
%\caption{Fidelity of composite pulse sequences for (a) the 3-pulse sequences
%(solid) and (b) the 5-pulse sequences (dashed), with coefficients for the
%6$^{th}$ order term of the fidelity given in Table 1. The fidelity of the
%single error-prone pulse is also shown (dotted).}
%\label{Fig4}
%\end{figure}

\end{document}